\documentclass[twocolumn,prb,aps,superscriptaddress,longbibliography]{revtex4-2}
\usepackage{amsmath}
\usepackage{amssymb}
\usepackage{amsfonts}
\usepackage[dvips]{graphicx}
\usepackage{subfigure}
\usepackage{dcolumn}
\usepackage{txfonts}
\usepackage{bm}
\usepackage{makeidx}
\usepackage{color}
\usepackage{mathtools}
\usepackage{threeparttable}
\usepackage{physics}
\usepackage[colorlinks,linkcolor=blue,anchorcolor=blue,citecolor=blue,urlcolor=blue]{hyperref}
\usepackage{lipsum}

\begin{document}
\title{Poor Man’s Majoranon in Two Quantum Dots Dressed by Superconducting Quasi-Excitations}

\author{Zhi-Lei Zhang}
\affiliation{Graduate School of China Academy of Engineering Physics, Beijing 100193, China}

\author{Guo-Jian Qiao}
\email{qiaoguojian19@gscaep.ac.cn}
\affiliation{Graduate School of China Academy of Engineering Physics, Beijing 100193, China}
\affiliation{Beijing Computational Science Research Center, Beijing 100193, China}
\author{C. P. Sun}
\email{suncp@gscaep.ac.cn}
\affiliation{Graduate School of China Academy of Engineering Physics, Beijing 100193, China}
 
\begin{abstract}
In a hybrid system consisting of two quantum dots (QDs) coupled to a superconductor (SC), zero-bias peaks in the differential conductance spectrum have been reported as potential signatures of Majorana fermions (MFs). However, such signatures typically appear only at specific parameter values of the QDs—so-called `sweet spots'—and are referred to as the Poor Man’s Majorana (PMM). To investigate whether these signatures can be conclusively attributed to genuine MFs emerging over a continuous parameter range, we present an alternative approach that microscopically incorporates the superconducting effects into the QDs, rather than simply attribute them into two phenomenological parameters of QDs. This forms the dressed Majorana fermions (DMFs), which can be viewed as superpositions of quasi-excitations from both the QDs and the SC. We show that DMFs can localize at one end of a one-dimensional SC and persist across a continuous parameter range, thereby enhancing the feasibility of experimental detection. Our results provide a more accurate description of the PMM in such hybrid systems and offer practical guidance for observing end-localized PMM modes in continuous one-dimensional SC.
\end{abstract}

\maketitle

\section{Introduction\label{sec: appendix1}}
Majorana fermions (MFs) in condensed matter systems, also referred to as Majorana zero modes, are fundamental to topological quantum computing~\cite{kitaev2001unpaired, PhysRevLett.104.040502, alicea2012new, lutchyn2018majorana, prada2020andreev}. Among the various proposed platforms for MFs, hybrid systems consisting of semiconducting nanowires coupled to $s$-wave superconductor (SC) have received widespread attention~\cite{PhysRevLett.105.077001, PhysRevLett.105.177002}. This is because the zero-bias peaks have been consistently reported in such systems, serving as experimental evidence supporting the existence of MFs~\cite{mourik2012signatures, Deng_2012,das2012zero, Nichele_2017, Zhang2018, Vaiti_2020, Zhang2022}. However, whether the observed zero-bias conductance peaks are indeed attributable to MFs remains a subject of intense debate~\cite{Zhang2018, Marco_2021_Nontopological_zero-bias, Frolov2021quantum, Zhang2022, Qiao_2022}. One of the primary challenges is that disorder and the inhomogeneous chemical potential in the hybrid nanowire system can also produce a similar zero-bias signature, thereby causing ambiguity in their interpretation~\cite{PhysRevB.86.100503, PhysRevB.96.075161, li2014probing, PhysRevResearch.2.013377, Sankar_Das_Sarma2023}. Indeed, many of the reported zero-bias signatures could likely originate from these sources rather than from MFs themselves~\cite{prada2020andreev, Sankar_Das_Sarma2023,kouwenhoven2025}.

In light of the aforementioned controversies, a simple hybrid system, which consists of two quantum dots (QDs) coupled to SC~\cite{sau2012realizing, Flensberg2012, fulga2013adaptive}, has been reconsidered for recent years~\cite{PhysRevB.110.115302, luethi2024perfect, fateofDPPM, Creating2022, liu2024enhancing}. This hybrid system can be theoretically modeled as a system of two spinless QDs where the phenomenological parameters for pairing and tunneling reflect the role of SC. By adjusting the magnetic field and the chemical potential in two QDs, the Majorana-like modes can emerge and be completely localized at the left and right two QDs (referred to as full polarization)~\cite{Flensberg2012}. However, this hybrid system has not garnered substantial attention because this MF can only manifest at certain well-defined parameter points, termed 'sweet spots', rather than over a region in parameter space~\cite{alicea2012new, Qiao_2022, Qiao2024}. In this sense, they were named as the `Poor Man's Majorana' (PMM). In fact, the realization of the PMM heavily depends on whether the ideal coupling between QDs through crossed Andreev reflection and elastic co-tunneling can be induced from a microscopic model. And the achievement of sweet spots also requires precise tuning of the strengths of them between the QDs, which is difficult to experimentally control at the time.

Recently, with advances in experimental techniques, zero-bias peaks in the differential conductance spectrum for such a hybrid system have been observed, as a possible signature of PMM ~\cite{dvir2023realization, ten2024two, zatelli2024robust, PhysRevX.13.031031} at the sweet spot. However, whether the zero-bias signatures are indeed caused by genuine MFs remains an open question because the experiments can not show the localizations of PMM (full polarization) in a QD explicitly. 

Especially, according to the most recent studies based on an effective theory of QDs by eliminating quasi-excitations in SC~\cite{Yue2023, Qiao_2022}, the PMMs indeed do not exist in the actual hybrid system~\cite{luethi2024perfect} with polarization. In fact, the effective theory is only applicable in the lower energy region, thus, it can only work well under conditions of low magnetic fields and weak tunneling strength between QDs and SC, i.e., Zeeman energy of QDs and tunneling strength are less than the gap of SC. In contrast, in practical experiments to realize the PMM, a strong magnetic field (the corresponding Zeeman energy $\sim 0.4\rm{meV}$ is large than the gap of SC ~\cite{dvir2023realization}) is often required to completely align all spins forming a spinless system, and the tunneling strength between the QDs and SC needs to be arbitrarily adjusted to reach the sweet spot~\cite{dvir2023realization, ten2024two}. 

Moreover, since the previous theory eliminated quasi-excitations in SC, it failed to characterize the spatial distribution of MFs and demonstrate the polarization of PMM— the end (edge) localization of PMM in 1D (or 2D) SC. Therefore, we need to propose an exactly solvable theory that treats QDs and SC on an equal footing — rather than eliminating the SC — to enable descriptions valid for arbitrary magnetic fields and tunneling strengths, while also revealing PMM localization dependent on the SC configuration.

To thoroughly address all the aforementioned theoretical and experimental challenges,  we develop a microscopic approach treating the SC and QDs on equal footing~\cite{Qiao2024} to analyze the emergence of PMM. Here, the superconducting segment is modeled as a one-dimensional (1D) long chain coupling to the two QDs, forming dressed quasi-excitations. Our analysis shows that PMMs emerge across a continuous parameter range, significantly improving experimental observability. Specifically, when the magnetic field is in the range of $-0.1$ to $-1 \rm{T}$ \cite{dvir2023realization}, PMMs can appear if the chemical potential of QDs is adjusted between $0$ and $3 \rm{meV}$, while the tunneling strength is correspondingly tuned to be within the range of $0.1$ to $40 \rm{meV}$. Moreover, we demonstrate that these PMMs localize at the SC ends, explicitly exhibiting particle-like characteristics. We term this state as `dressed poor man’s Majorana' (DPMM), which persists for arbitrary tunneling strengths. Additionally, our approach provides a microscopic description of polarization of PMM~\cite{Creating2022, liu2024enhancing, luethi2024perfect}, herein defined by the wave function of PMM localized at the two ends of the 1D SC.

Additionally, we show that the wave function distribution of DPMM is strongly influenced by the tunneling strength. As the strength increases, the DPMM wave function gradually transfers from the QDs to the SC. Specifically, when tunneling between the QDs and SC is sufficiently strong, the spatial distribution of PMM within the SC will dominate over that within the QDs. This result highlights that explicit consideration of the SC’s role is essential to determine whether PMM is full polarization (i.e., localized at the ends). Moreover, we demonstrate that under weak tunneling and strong magnetic fields, the DPMM state reduces to the PMM state described by the original phenomenological model~\cite{Flensberg2012}. This indicates that our QDs-SC model degenerates into the phenomenological framework under these conditions.

\section{Dressed poor man's Majoranons in microscopic model\label{sec: appendix2}}
\begin{figure}
    \centering
    \includegraphics[width=8cm]{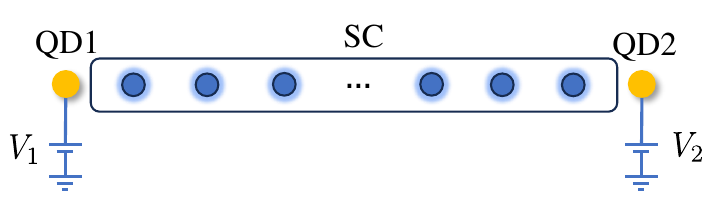}   \caption{Sketch of the hybrid system where the SC is described by a 1D lattice model, and the two QDs are coupled to the first and last lattice sites of the SC, respectively. $V_i\;(i=1,2)$ is the gate voltage to adjust the chemical potential of the QDs.} \label{sketch map}
\end{figure}

In this section, we revisit the microscopic model of the two quantum dots (QDs) - superconductor (SC) hybrid system. Then we define the dressed poor man's Majoranons (DPMMs) by considering both the QDs and the quasi-excitation of SC, and further determine the condition for its existence. The Hamiltonian of two QDs hybrid system \cite{luethi2024perfect, liu2024enhancing, sau2012realizing, Qiao_2025} is $H = H_d + H_s + H_t$, with the Hamiltonian of two QDs: 
\begin{equation}
    H_{d} = \sum_{i=1,2} (\mu_d + h_d) d_{i \uparrow}^{\dagger} d_{i \uparrow} + (\mu_d - h_d) d_{i \downarrow}^{\dagger} d_{i \downarrow},
\end{equation}
where the intra-dot Coulomb interactions are neglected for simplicity~\cite{Yeyati1997, sun1999, Qiao_2025}, an $s$-wave SC is desicrbed by 1D lattice model as \cite{Qiao2024}:
\begin{equation}
H_s = \sum_{n,\sigma}[\mu_{s}c_{n,\sigma}^{\dagger}c_{n,\sigma}-\frac{1}{2}(t_{s}c_{n,\sigma}^{\dagger}c_{n+1,\sigma}+\Delta_{s}c_{n\uparrow}^{\dagger}c_{n\downarrow}^{\dagger} + {\rm{h.c.}})],
\end{equation}
and tunneling Hamiltonian between QDs and SC is~\cite{fateofDPPM, liu2024enhancing, Creating2022, luethi2024perfect,Qiao_2025}
\begin{equation}
\begin{aligned}
 H_{t} &= \sum_{\sigma} T d_{1 \sigma}^{\dagger} c_{1 \sigma} + \alpha [d_{1 \uparrow}^{\dagger} c_{1 \downarrow} - d_{1 \downarrow}^{\dagger} c_{1 \uparrow}] + {\rm h.c.} \\
    &+ \sum_{\sigma} T d_{2 \sigma}^{\dagger} c_{N \sigma} + \alpha [d_{2 \downarrow}^{\dagger} c_{N \uparrow} - d_{2 \uparrow}^{\dagger} c_{N \downarrow}] + {\rm h.c.}\\
\end{aligned}
\end{equation}
Here, $c_{n\sigma} (c_{n\sigma}^{\dagger})$ and $d_{i\sigma} (d_{i\sigma}^{\dagger})$ represent the annihilation (creation) operator of the SC and QD with spin $\sigma$ on site $n$ or $i$, respectively. Moreover, $\Delta_s$ is the pairing strength, $t_s$ is the hopping strength of the adjacent lattice sites in the SC, $\mu_d$ and $h_d$ are the on-site energy and the Zeeman energy of two QDs, $T$ is the strength of tunneling between the QDs and SC, $\alpha$ is the spin-orbit coupling strength. The chemical potential of QDs $\mu_d$ can be adjusted experimentally by the applied gate voltage $V_i$, as shown in Fig. \ref{sketch map}. 

In the earliest studies, this QD-SC hybrid system was modeled as two coupled spinless QDs, which predicted the existence of Poor Man's Majorana (PMM)~\cite{Flensberg2012}. In this simplified description, the SC was oversimplified and treated merely as a medium to provide two phenomenologically coupled parameters between QDs. By eliminating excitations in the SC, recent theory~\cite{luethi2024perfect} demonstrates that this treatment is inadequate, and PMM fully localized within QDs does not exist. However, because the superconducting quasi-exitations have been neglected, this theory is only applicable at low energy scales and cannot characterize the spatial distribution of PMM in the SC. Therefore, we need to consider a microscopic theory that incorporates both electrons in QDs and excitations in SC, as discussed below.

We diagonalize the Hamiltonian of the hybrid system as $H = \sum_\nu E_{\nu} \gamma_{\nu}^{\dagger} \gamma_{\nu}$, where  
\begin{equation}
    \gamma_{\nu} = \mathbf{u}_\nu^d \cdot \mathbf{d} + \mathbf{d}^{\dagger} \cdot (\mathbf{v}_\nu^d)^T + \mathbf{u}_\nu^s \cdot \mathbf{c} + \mathbf{c}^{\dagger} \cdot (\mathbf{v}_\nu^s)^T \label{quasi-particle operation}
\end{equation}
with $\mathbf{d} = [d_{1\uparrow}, d_{1\downarrow}, d_{2 \uparrow}, d_{2\downarrow}]^T$ and $\mathbf{c}=[c_{1\uparrow},c_{1\downarrow},\ldots,c_{N\uparrow},c_{N\downarrow}]^{T}$. $\gamma_{\nu}$ is the annihilation operator of quasi-particle operation, and 
\begin{equation}
\begin{aligned}
\mathbf{v}_\nu^\alpha &= [v_{\nu,1\uparrow}^\alpha, v_{\nu,1\downarrow}^\alpha, \dots, v_{\nu, N_\alpha\uparrow}^\alpha, v_{\nu, N_\alpha\downarrow}^\alpha],\\
\mathbf{u}_\nu^\alpha &= [u_{\nu,1\uparrow}^\alpha, u_{\nu,1\downarrow}^\alpha, \dots, u_{\nu, N_\alpha\uparrow}^\alpha, u_{\nu, N_\alpha\downarrow}^\alpha],
\end{aligned}
\end{equation}
with $\alpha=d,s$ are the electron and hole wave function for SC ($\alpha=s, N_s=N$) and QD ($\alpha=d, N_d=2$).

The DPMM can be defined by $\gamma_\nu = \gamma_\nu^{\dagger}$, similar to the dressed Majorana fermions defined in Ref. \cite{Qiao2024}. In this way, the electron and hole wave functions satisfy $\mathbf{u}_\nu^o = (\mathbf{v}_\nu^o)^*, (o = d,s)$. Due to the particle-hole symmetry of the system \cite{li2014probing, Qiao_2022, Qiao2024}, the eigen mode of the DPMM is zero, i.e., $E_{\nu} = 0$. Consequently, we can establish the condition for the existence of DPMM by solving the zero-energy eigen equation (see Appendix \ref{appendixA}). The derived condition for the presence of DPMM is
\begin{equation}
    \left[\mu_d - \Re(\chi)\right]^2 + \Im(\chi)^2 = h_d^2 \label{sweet spot}
\end{equation}
with 
\begin{equation}
    \begin{aligned}
        \chi = \frac{2(T^2 + \alpha^2) \xi}{t_s},\;\xi =\frac{\mu_s - i \Delta_s - \sqrt{(\mu_s - i \Delta_s)^2 - t_s^2}}{t_s}.
    \end{aligned}
\end{equation}
Eq. \eqref{sweet spot} determines the range of parameters for the existence of DPMM and is suitable for arbitrary magnetic field and tunneling strength as long as these parameters do not disrupt the structures of SC or QD. Moreover, it is also seen from Eq. \eqref{sweet spot} that $\Re(\chi)$ represents the shift of the chemical potential of the QD and $\Im(\chi)$ is akin to the induced effective gap in QDs. They are induced by the dressed effect of SC. When this effect becomes strong, the shift of the chemical potential becomes very apparent, and the effective gap will become increasingly larger.

\begin{figure}
    \centering
    \includegraphics[width=8.5cm]{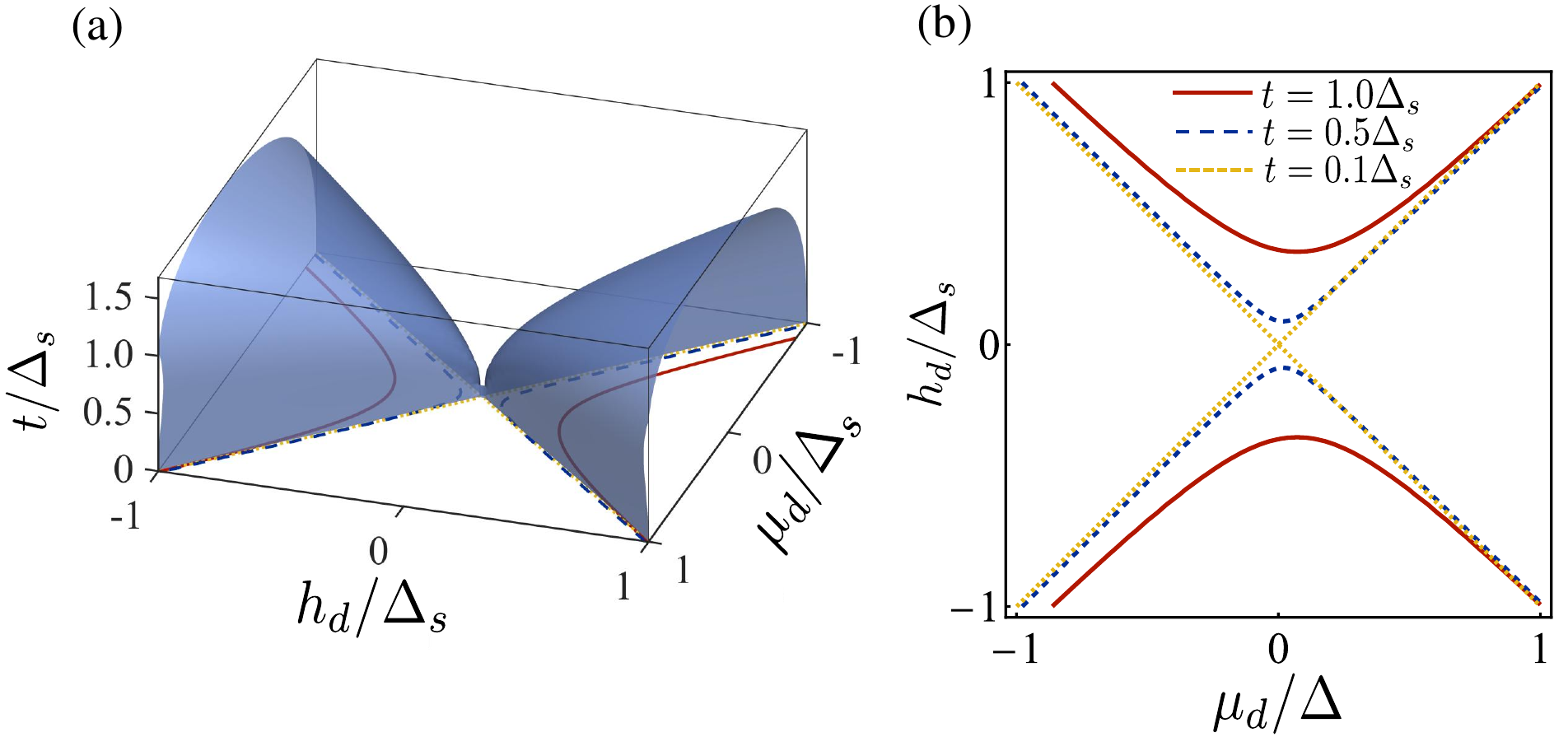}   \caption{(a) DPMM exists on the surface of $h_d-\mu_d-t$ space, where $T=\alpha \equiv t$. (b) The existence condition for the DPMM in different tunneling strengths: the weak tunneling at $t=0.1\Delta_s$ (the yellow dashed line), the intermediate tunneling at $t=0.5\Delta_s$ (the blue dotted line), and the strong tunneling at $t=1.0\Delta_s$ (the red solid line). The parameters of the SC are set as $\mu_s=2\Delta_s$ and $t_s=10\Delta_s$.} \label{phase diagramTh}
\end{figure}

\begin{figure}
    \centering
    \includegraphics[width=8.5cm]{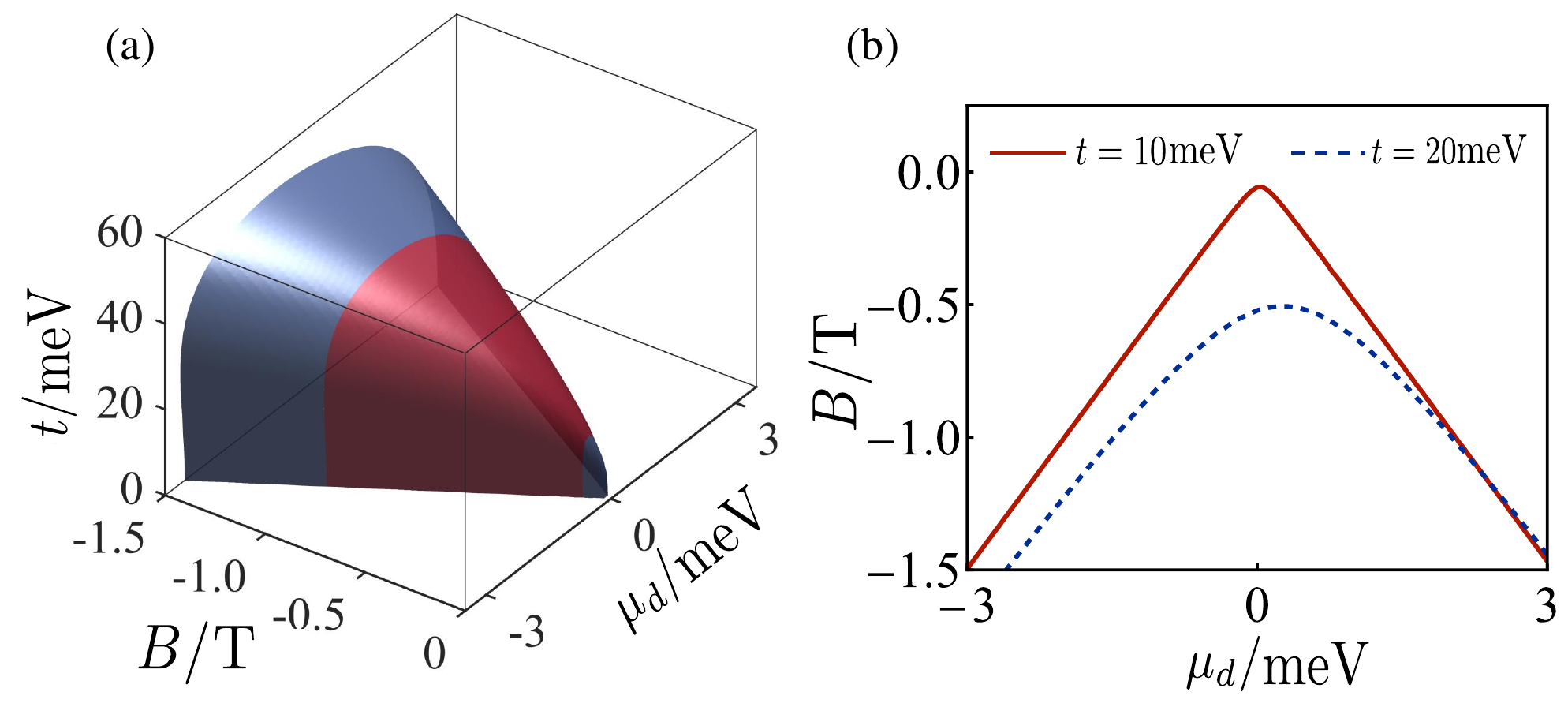}   \caption{Condition for the existence of DPMM in the practical experimental parameters $g=35, \Delta_s \approx 0.34{\rm{meV}}, \mu_s = 2.5\times10^3 \Delta_s$ and $t_s = 10^4 \Delta_s$, (a) DPMM exists on the surface of $B-\mu_d-t$ space, where $B$ is the magnetic field applied in the QDs. When the tunneling strength is continuously adjusted, DPMM can emerge within a continuous region, where the magnetic field ranges from $0.1-1\mathrm{T}$, and the chemical potential of QDs is between $0-3 \rm{meV}$ (red surface).  (b) For a given tunneling strength, DPMM exhibits in the continuous range of parameters in the $B-\mu_d$ space.} \label{phase diagramEx}
\end{figure}

To clearly show these results, Fig. \ref{phase diagramTh}(a) plots the surface for the existence of DPMM in the space of chemical potential, Zeeman energy, and tunneling strengths, denoted by $T = \alpha \equiv t$. As the tunneling strength increases from weak tunneling $t = 0.1\Delta_s$ (blue dotted line) to stronger tunneling $t = 0.5\Delta_s$ (yellow dashed line) and $t = \Delta_s$ (red solid line), the shift of the chemical potential of the QDs, and the effective gap of the QDs become larger, as observed in Fig. \ref{phase diagramTh} (b). As a result, DPMMs as well as PMMs appear across the continuous parameter range, rather than at a specific point. This feature makes experimental observations more feasible. For example, in the practical QDs-SC platform~\cite{dvir2023realization}, the QDs is formed by the $\rm{InSb}$ nanowire, and $\rm{Al/Pt}$ is used as a superconducting component. Under these parameters for the practical material, where Land\'e factor $g=35,\Delta_s \approx 0.34\,{\rm{meV}}, \mu_s = 2.5\times10^3 \Delta_s$ and $t_s = 10^4 \Delta_s$, DPMM can exist across the continuous parameter within the magnitude of magnetic field $\sim0.1-1\rm{T}$ and $|\mu_d | \sim 0-3\rm{meV}$ when the tunneling strength is continuously adjusted from $0.1 \rm{meV}$ to $40 \rm{meV}$, as shown in the red region of Fig. \ref{phase diagramEx} (a). Moreover, for a given tunneling strength shown in Fig. \ref{phase diagramEx} (b), DPMMs emerge when the chemical potential increases from 0 to $3 \rm{meV}$, while the magnetic field is correspondingly adjusted from $-0.1$ to $-1.5 \rm{T}$. Then, we need to determine whether the DPMMs are spatially localized, which is a characteristic feature of Majorana fermions as particles.

\section{The spatial distribution of Dressed Poor man's Majoranon}
In this section, we present the spatial distribution of the wave function of DPMMs, further demonstrating that DPMMs exhibit particle-like characteristics. This is the reason why we call them as dressed poor man's `Majoranons'. We also microscopically clarify what `full polarization' means for PMM based on the DPMMs' picture. 

By solving the zero-energy wave function, the DPMMs are obtained as  
\begin{equation}
\begin{aligned}
    \gamma_{\nu} &= \mathbf{u}_\nu^d \cdot \mathbf{d} + [\mathbf{u}_\nu^d \cdot \mathbf{d}]^{\dagger} + \mathbf{u}_\nu^s \cdot \mathbf{c} + [\mathbf{u}_\nu^s \cdot \mathbf{c}]^{\dagger}.
    \end{aligned} 
    \label{realimagegamma}
\end{equation}
These DPMMs are a composite of electron and hole excitation in QD dressed by the quasi-excitation of SC [the last two terms of Eq. \eqref{realimagegamma}], and it differs from the PMM defined in two QDs by the phenomenological or low-energy theory, where the quasi-excitations in SC have been eliminated. Here, the wave function of the electron within SC is (see Appendix \ref{appendixA})
\begin{equation}
\begin{aligned}
    \textbf{u}_{1,n}^s &= \sum_{l=1,2} \alpha_l \xi_{l}^n \mathbf{a}_l^{(1)},\;\;\textbf{u}_{3,n}^s = i \sum_{l=3,4} \alpha_l \xi_{l}^{n+1-N} \mathbf{a}_l^{(2)}\\
    \textbf{u}_{2,n}^s &= i \sum_{l=1,2} \beta_l \xi_{l}^{n} \mathbf{a}_l^{(2)},\;\;\textbf{u}_{4,n}^s = \sum_{l=3,4} \beta_l \xi_{l}^{n+1-N} \mathbf{a}_l^{(1)}
    \end{aligned} \label{wave function of SC}
\end{equation}
where $1 \leq n \leq N$. $\xi_1 = \xi$ and $\xi_2 = \xi^{*}$ with $\abs{\xi_{1(2)}}<1$ is the solved decay factor of the wave function of the DPMM. In contrast, $\xi_{3(4)} = \xi_{1(2)}^{-1}$ are gain factor, and $\abs{\xi_{3(4)}}>1$.  $\mathbf{a}_l = [a_{l\uparrow}, a_{l\downarrow}]^T$ is the eigenvector of the zero-energy eigen-equation for a given $\xi_l$. Moreover, $\alpha_l,\beta_l$ are superposition coefficients determined by the boundary and normalization conditions (see Appendix \ref{appendixA}).
\begin{figure}
    \centering
    \includegraphics[width=8.5cm]{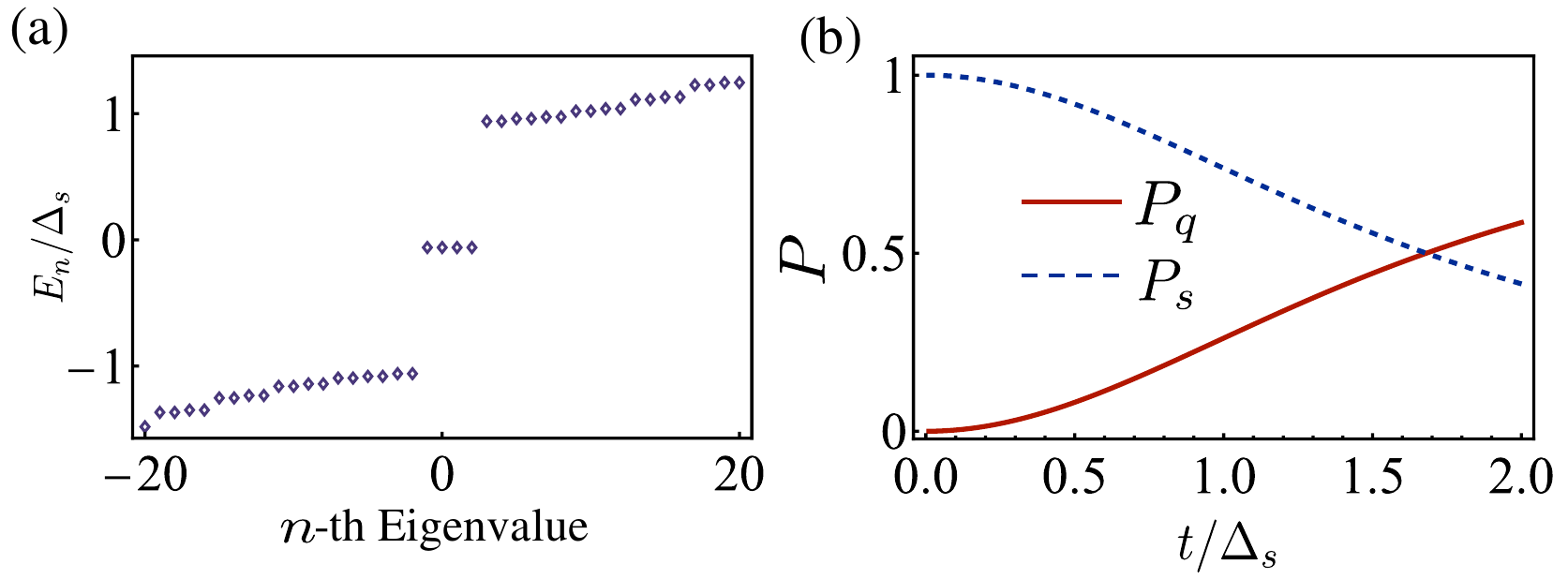}   \caption{(a) The eigenvalue of the QDs-SC system. (b) The weight of zero-energy wavefunctions distributed in SC and QDs with the increasing tunneling strength given $T = \alpha \equiv t$. The parameters are set as $N=150, \mu_s=2\Delta_s,t_s = 10\Delta_s,T=\alpha = 1.5\Delta_s,\mu_d = 0.7\Delta_s$ and $h_d \approx 0.961\Delta_s$.} \label{eigenvalue}
\end{figure}
Based on Eqs (\ref{realimagegamma}, \ref{wave function of SC}), it is easy to demonstrate that $\mathbf{u}_1^s$ and $\mathbf{u}_2^s$ are localized at one end of the SC due to $\abs{\xi_{1(2)}}<1$, while $\mathbf{u}_3^s$ and $\mathbf{u}_4^s$ are localized at the opposite end of the SC since  $\abs{\xi_{3(4)}}>1$. This reflects the characteristics of edge states. Additionally, we have numerically verified that there exist four DPMMs when Eq. \eqref{sweet spot} is satisfied [see Fig. \ref{eigenvalue} (a)], and the spatial distribution of these edge states in SC is indeed localized, see Fig. \ref{wavefunction in SC} (a-d). 

\begin{figure}
    \centering
    \includegraphics[width=8.5cm]{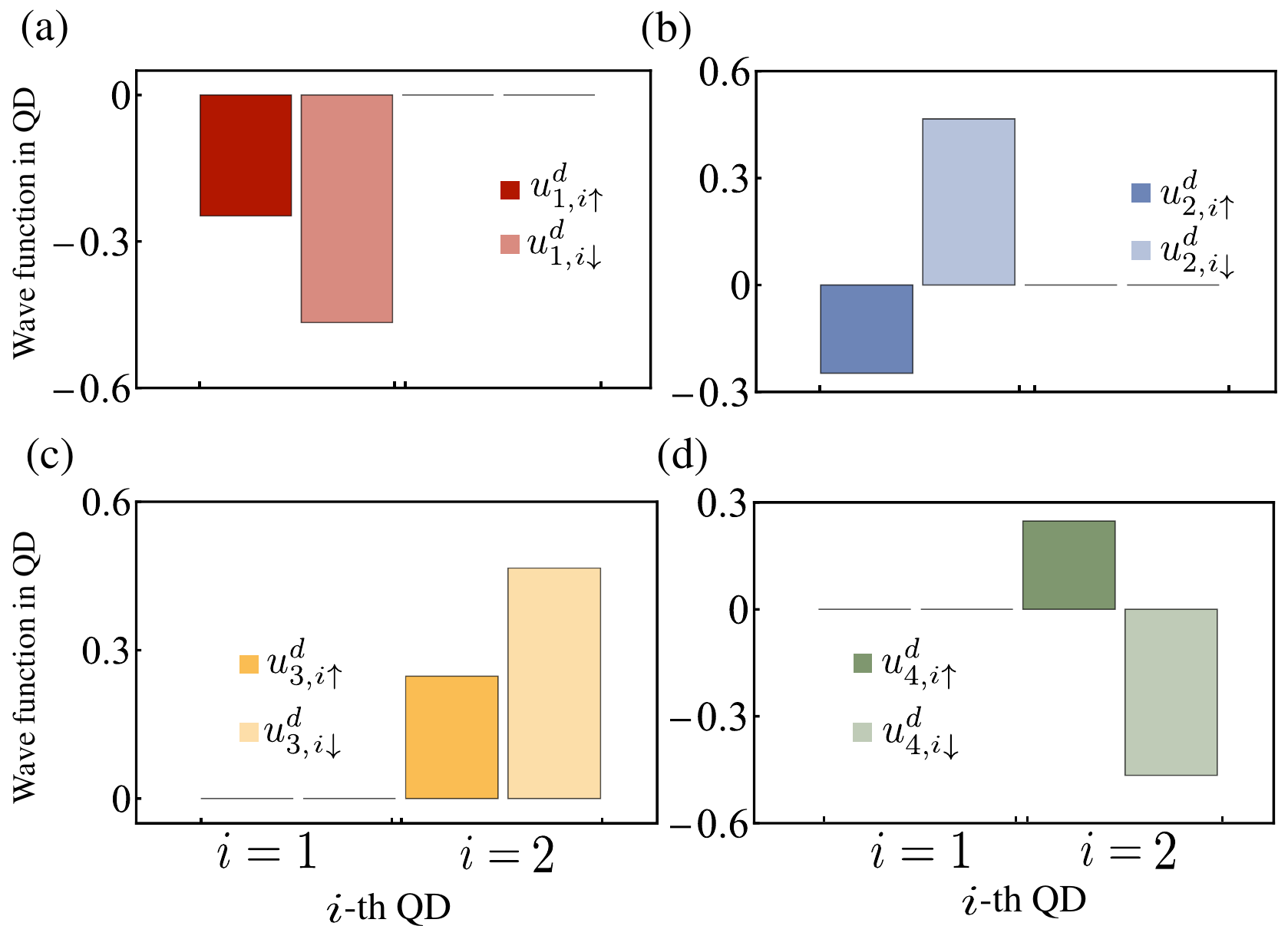}   \caption{The spatial distribution of zero-energy wave function at $i$-th QD with $\sigma$. The parameters are taken same as Fig. \ref{eigenvalue}.} \label{wavefunction in QD}
\end{figure}

Furthermore, the wave function in the QDs is given by $\mathbf{u}^{d}_{\nu} = - \mathbf{H}_d^{-1} \cdot \mathbf{T} \cdot \mathbf{u}^{s}_\nu$, where $\nu=1,2,3,4$. The distribution of these wave functions within the two QDs is also depicted in Fig. \ref{wavefunction in QD}, where the height of the bar chart represents the spatial distribution of the wave function of $\nu$-th DPMM on the $i$-th QD with spin $\sigma$. Specifically, $\mathbf{u}_{1,2}^d$ represents the wave function of the zero energy localized separately at the left QD, while $\mathbf{u}_{3,4}^d$ corresponds to those localized at the right QD. Considering the spatial distribution of the wave functions of DPMM in both SC and QDs, we have proven that DPMM exhibits Majorana particle-like characteristics. This also microscopically defines the `full polarization'. Namely, the wave function of DPMM is localized at both ends of the SC with QDs. Importantly, the spatial distributions of the two DPMM wave functions do not overlap. 

Moreover, the distribution weight of zero-energy wave functions in SC and QDs can be characterized by $P_{\alpha} = \sum_{n\sigma} |\mathbf{u}_{n\sigma}^{\alpha}|^2 + |\mathbf{v}_{n\sigma}^{\alpha}|^2$. In Fig. \ref{eigenvalue} (b), we illustrate how the distribution weight varies as the tunneling strength increases. With an increase in tunneling strength (giving $T = \alpha \equiv t$), a larger weight of the zero-energy wave function extends from the QDs into the SC, which is also shown in other hybrid systems~\cite{Qiao2024, liu2024enhancing}. When the tunneling strength is moderate or even quite strong (i.e., $t$ is greater than $\Delta_s$), the zero-energy wave function within the SC becomes comparable to that in the QDs (for instance, when $t$ is approximately $1.7 \Delta_s$ and the wave function weight in SC and QD are both around $50 \%$). As a result, the effects of the SC become significant and cannot be ignored. 

\begin{figure}
    \centering
    \includegraphics[width=8.5cm]{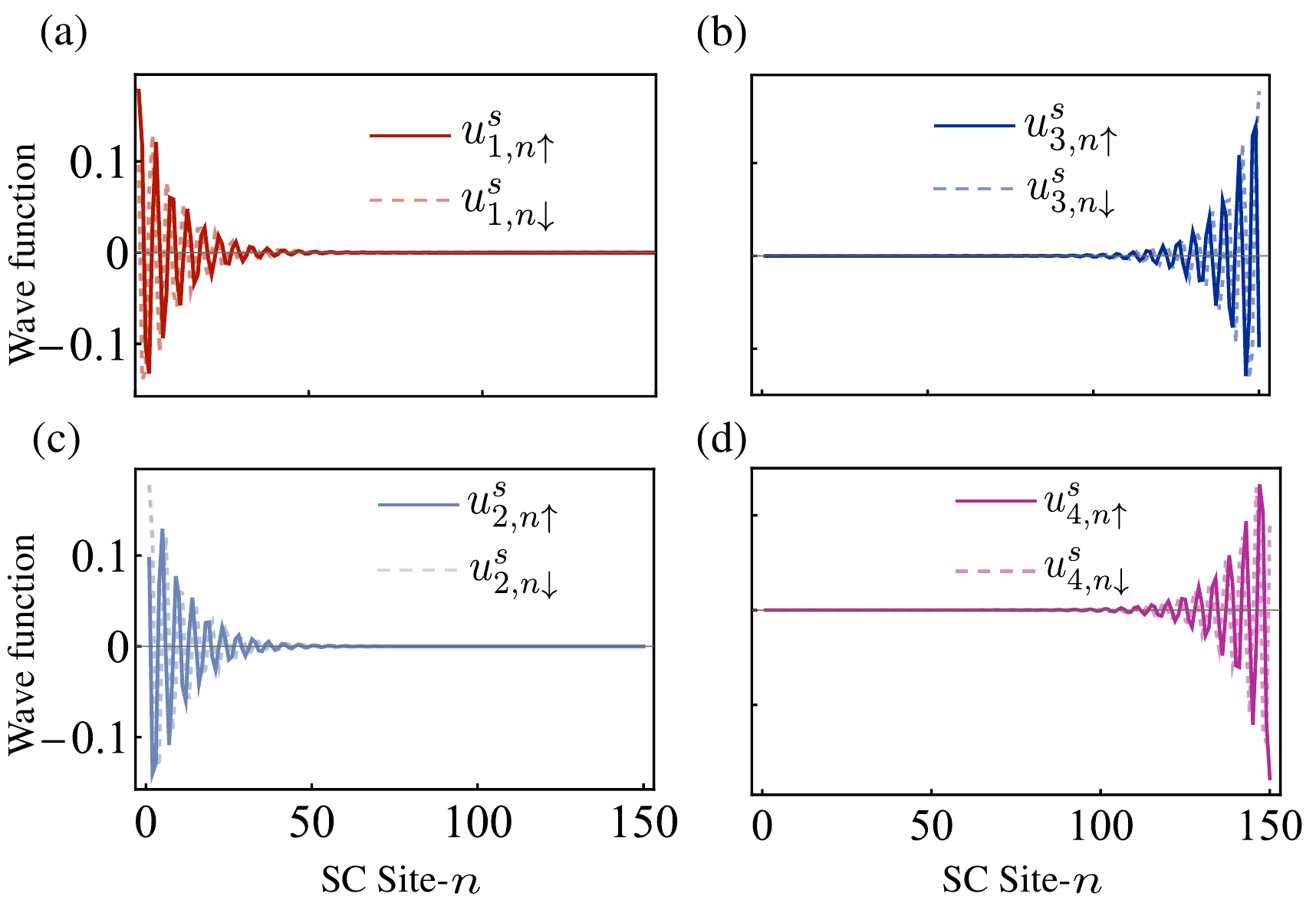}   \caption{The spatial distribution of zero-energy wave function at each site $n$ in SC. The parameters are taken same as Fig. \ref{eigenvalue}.} \label{wavefunction in SC}
\end{figure}

\section{The Condition of Spin Polarization and the Applicability of the phenomenological Model}
When the applied magnetic field is sufficiently strong, the electron spins in QDs become polarized, aligning either in the upward or downward direction. This polarization is a necessary condition for realizing the phenomenological model of PMM~\cite{Flensberg2012}. However, it is unclear under which practical physical conditions the hybrid system can be characterized by the phenomenological model, i.e., the condition of spin polarization. In this section, we employ the fidelity $F$ to assess the `distance' between the DPMM state and the PMM state defined in the phenomenological model (ideal PMM state)~\cite{Flensberg2012}. $F = 1$ indicates that the DPMM and PMM are identical, and it determines the physical conditions under which the phenomenological model is valid. When $F = 0$, the two quantum states are orthogonal. In this case, spin polarization cannot be achieved, causing the DPMM in the QD-SC hybrid system to deviate significantly from the ideal PMM state in the phenomenological model. At this point, the phenomenological model is no longer suitable for describing the hybrid system.

The DPMM state is defined in the single excitation subspace by
\begin{equation}
    \ket{\psi_\nu} = \gamma_\nu^{\dagger} \ket{{\rm{vac}}},
    \label{single_occupy_state}
\end{equation}
where $\gamma_\nu^{\dagger}$ is the operator of DPMM, has been given in Eq. \eqref{realimagegamma}, and $\ket{\rm{vac}} = \ket{0_d} \otimes\ket{0_s}$ is the vacuum state, that is, there are no electrons occupying any sites in both SC and QDs. By Eq. \eqref{single_occupy_state}, the density matrix of the hybrid QDs system is given by 
\begin{equation}
    \rho_\nu = \ket{\psi_\nu}\bra{\psi_\nu}.
\end{equation}
Further, the reduced density matrix of the QDs is
\begin{equation}
    \begin{aligned}
        \rho_{\nu,d} = \Tr_s (\rho_\nu) = \sum_{n_s} \bra{n_s} \gamma_\nu^{\dagger} \ket{\rm{vac}} \bra{\rm{vac}} \gamma_\nu \ket{n_s},
    \end{aligned}
\end{equation}
where $\ket{n_s}$ represents the state with $n_s$ electron excitations: 
\begin{equation}
     \ket{0_s},\;\ket{1_s} =c_{m,\uparrow}^{\dagger}\ket{0_s}, c_{m,\downarrow}^{\dagger}\ket{0_s},(m=1,\dots,N)\dots
\end{equation}
Therefore, the reduced density matrix of the QDs is reduced to 
\begin{equation}
\begin{aligned}
    \rho_{\nu,d} &= \sum_{i,j}\sum_{\sigma = \uparrow,\downarrow}\sum_{\sigma^{\prime} = \uparrow,\downarrow} u_{\nu,i\sigma}^d (u_{\nu,j\sigma^{\prime}}^d)^* \ket{i_{\sigma,d}} \bra{j_{\sigma^{\prime},d}} \\
    &+ \sum_i \left(\abs{u_{\nu,i\uparrow}^s}^2 + \abs{u_{\nu,i\downarrow}^s}^2\right) \ket{0_d}\bra{0_d},
\end{aligned}\label{RDM of DPMM}
\end{equation}
where $\ket{i_{\sigma,d}}=d_{i,\sigma}^{\dagger}|0_{d}\rangle$ represents the state that one electron with spin $\sigma$ occupies the $i$-th QD. Note that the proportion of QD with zero occupation is determined by the distribution of the zero-energy wave function in the SC, which manifests the dressing effect from the SC.

\begin{figure}
    \centering
    \includegraphics[width=8.5cm]{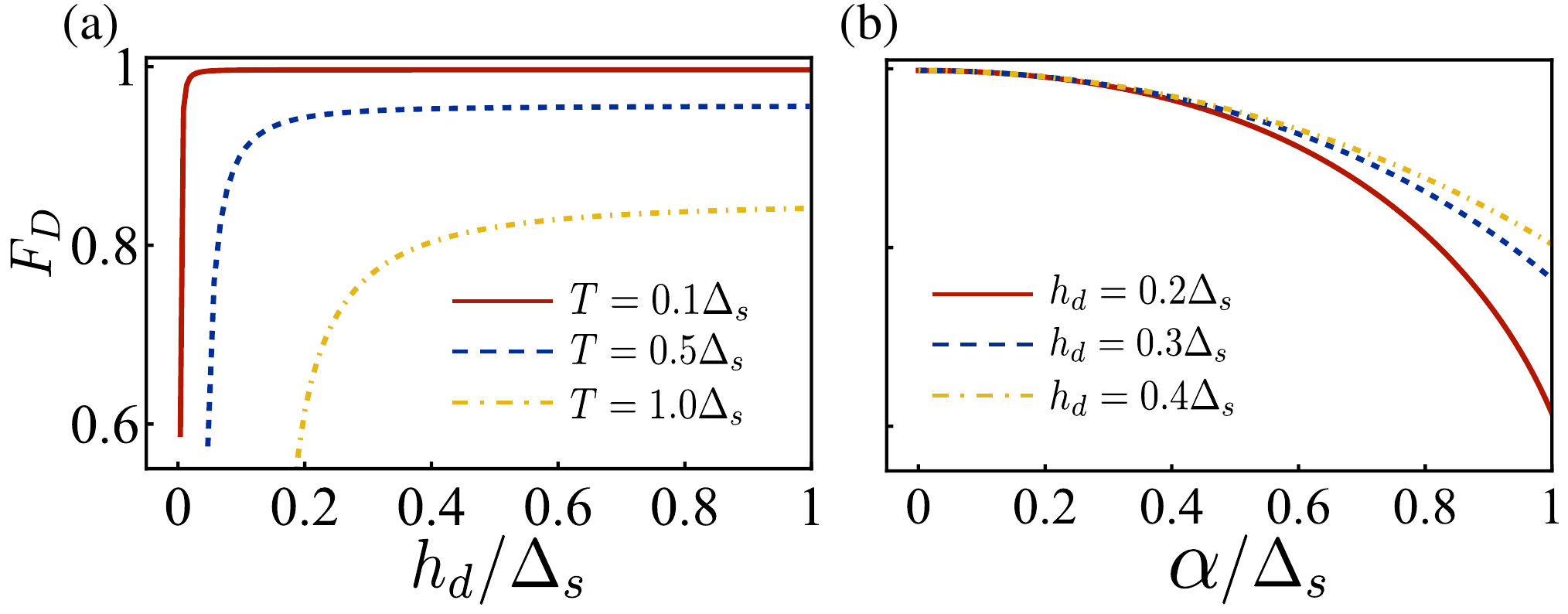}   \caption{(a) Fidelity between the DPMM and the PMM with respect to the Zeeman energy $h_d$ in different tunneling strengths $T$ with $\alpha = 0.1 \Delta_s$. (b) The dependence of the fidelity on the spin-orbit coupling $\alpha$ is plotted under different Zeeman energy $h_d$, with $T = 0.1 \Delta_s$. The other parameters are set as $N=150,\; \mu_s=2\Delta_s,\;t_s = 10\Delta_s$.} \label{Fidelity}
\end{figure}

Additionally, in hybrid QDs system with spin polarized in the down direction by an external magnetic field, the ideal PMM state should be
\begin{equation}
\begin{aligned}
    \ket{\varphi_1} &= (d_{1\downarrow}^{\dagger} + d_{1\downarrow}) \ket{0_d}= d_{1\downarrow}^{\dagger}\ket{0_d},\\
    \ket{\varphi_2} &= i (d_{2\downarrow}^{\dagger} - d_{2\downarrow}) \ket{0_d}= i \; d_{2\downarrow}^{\dagger}\ket{0_d},
\end{aligned}
\end{equation} 
which is the actual implementation of the PMM state in the phenomenological model. Then the fidelity of the DPMM and the PMM is defined by 
\begin{equation}
\begin{aligned}
    F_{1(2),D} &= \bra{\varphi_1} \rho_{1(2),d} \ket{\varphi_1},\\
    F_{3(4),D} &= \bra{\varphi_2} \rho_{3(4),d} \ket{\varphi_2}. \label{Fidelity of DPMM}
\end{aligned}
\end{equation}
By Eq. \eqref{Fidelity of DPMM}, we numerically calculated the fidelity of the DPMM and PMM states, and verified that these $F_{i,D}$ are equal, i.e., $F_{i,D}=F_{j,D} \equiv F_{D}$. It is seen from  Fig. \ref{Fidelity} (a) that for a given tunneling strength, i.e., $t=0.1\Delta_s$, the fidelity is a monotonically increasing function of the Zeeman energy. This indicates that the electron in QDs is gradually polarized towards the downward direction, thereby reaching the ideal PMM state. As the coupling strength increases, the polarization tendency gradually diminishes [see Fig. \ref{Fidelity}(a)]. This is because the weight of the zero-energy wave function in SC gradually becomes larger [see Fig. \ref{eigenvalue}(b)]. Fig. \ref{Fidelity} (b) shows that the Fidelity is a monotonically decreasing function with respect to the strength of spin-orbit coupling $\alpha$. This phenomenon can be attributed to two factors: (i) Spin-orbit coupling results from the interaction between spin-up and spin-down states, which weakens spin polarization as it becomes stronger. To preserve spin polarization, a stronger magnetic field is required. (ii) The enhancement of spin-orbit coupling will increase the wave function of DPMM within the SC, which in turn leads to a gradual increase in the zero occupancy of QDs. This also results in a reduction of fidelity. Therefore, spin polarization can only be achieved with weak tunneling and strong Zeeman energy. This further emphasizes that the dressing effect of SC on QDs under these regimes cannot be ignored.

\section{Conclusion}

We have investigated the influence of quasi-excitations in the superconductor (SC) on coupled quantum dots (QDs), demonstrating that the Poor Man's Majoranas (PMMs) is reduced from the so-called dressed Poor Man's Majoranons (DPMMs) composed of excitations in both QD and SC. Under the conditions of arbitrary magnetic fields and tunneling strengths, as long as they do not disrupt the superconducting structures and QD, PMMs as well as DPMMs can emerge in realistic QDs-SC hybrid systems. Significantly, the PMM appears within a continuous range of parameters rather than at specific points, such as the sweet spot ~\cite{Flensberg2012}. This feature makes the PMM more experimentally observable. Specifically, for the magnitude of magnetic field within $0.1 \sim 1\rm{T}$ \cite{dvir2023realization}, DPMM can emerge in the chemical potential of QDs between $0$ and $3\rm{meV}$ for a suitable tunneling strength ($0.1\sim40\rm{meV}$). With the picture of DPMMs, We also show that the PMMs are localized at the ends of the SC  with QDs, providing a microscopic description of the `full polarization' ~\cite{Creating2022, liu2024enhancing, luethi2024perfect}.

As found in our study, when the tunneling strength between QDs and SC increases, the proportion of the DPMM's wave function distributed within the SC becomes larger, and the full polarization of PMM with respect to the SC can be clearly presented. Therefore, the SC configuration is a crucial factor that cannot be ignored in the description of the PMM polarization problem. Additionally, we emphasize that in the case of weak tunneling and strong magnetic fields, the DPMM state will be reduced to the PMM state in the original phenomenological model ~\cite{Flensberg2012}, implying that the phenomenological model effectively describes the practical QDs-SC model under these conditions. Finally, we would like to stress that, although the SC is regarded as an infinite one-dimensional chain in the present approach, this method is also applicable to SC with finite length and transverse sizes.

\section*{acknowledgment \label{sec7}}

The authors would like to thank Xin Yue at CSRC for his valuable discussions. This work was supported by the National Natural Science Foundation of China (NSFC) (Grant No. 12088101) and NSAF No. U2330401.

\twocolumngrid
\appendix   
\setcounter{table}{0}   
\setcounter{figure}{0}
\renewcommand{\thetable}{A\arabic{table}}
\renewcommand{\thefigure}{A\arabic{figure}}

\section{The edge state of the dressed Poor man's Majoranon \label{appendixA}}

In this subsection, we present the conditions for the existence of the dressed Poor man's Majoranon (DPMM) and the wave function of these DPMMs. For the two quantum dots (QDs) coupled to the superconductor (SC), its Hamiltonian can be rewritten as
\begin{equation}
    \begin{aligned}
        H = \frac12 \mathbf{C}^{\dagger} \cdot \mathbf{H} \cdot \mathbf{C},\quad  \mathbf{H}=\begin{bmatrix}\mathbf{h} & \mathbf{p}\\
\mathbf{p^{\dagger}} & -\mathbf{h}
\end{bmatrix}.
\end{aligned}
\end{equation} 
Here, the vector operator is $\mathbf{C}^{\dagger} = [\mathbf{d}^{\dagger}, \mathbf{c}^{\dagger}, \mathbf{d}, \mathbf{c}]$ with $\mathbf{d} = [d_{1\uparrow}, d_{1\downarrow}, d_{2 \uparrow}, d_{2\downarrow}]^T$ and $\mathbf{c}=[c_{1\uparrow},c_{1\downarrow},\ldots,c_{N\uparrow},c_{N\downarrow}]^{T}$, $\mathbf{h}$ and $\mathbf{p}$ are $(2N+4)\times(2N+4)$ matrices
\begin{equation}
    \mathbf{h}=\begin{bmatrix}\mathbf{H}_{d} & \mathbf{T}\\
\mathbf{T}^{\dagger} & \mathbf{H}_{s}
\end{bmatrix},\quad\mathbf{p}=\begin{bmatrix}\mathbf{0} & \mathbf{0}\\
\mathbf{0} & \mathbf{\Delta}
\end{bmatrix},
\end{equation}
where the Hamiltonian matrices of the QDs, SC without the pairing term, and the tunneling term between the QDs and SC are
\begin{equation}
\begin{aligned}
\mathbf{H}_d &= (\mu_d \sigma_0 + h_d \sigma_z) \otimes \sigma_0\\
\mathbf{T} &= \begin{bmatrix}T & \alpha &0 &\dots & 0 &0\\
- \alpha & T &0 & \dots & 0 & 0\\
0& 0 & 0 & \dots & T & - \alpha\\
0 & 0 &0 & \dots &\alpha & T 
\end{bmatrix}_{4\times 2N}\\
\mathbf{H}_{s}&=\mu_{s}\sigma_{0}\otimes\tau_{0}-\frac{t_{s}}{2}\sigma_{0}\otimes\tau_{1},
\end{aligned}
\end{equation}
and the pairing term reads
\begin{equation}
\mathbf{\Delta}= - i\Delta_{s}\sigma_{y}\otimes\tau_{0}.
\end{equation}
Here, $\sigma_0 = {\rm{diag}}(1,1)$ is identity matrix, $\sigma_{x,y,z}$ are the Pauli matrices, $\tau_0 =\mathbb{I}_{N \times N}$ is the $N \times N$ identity, and 
\begin{equation}
    \tau_1 = \begin{bmatrix}
0 & 1 &  \cdots & 0 &0 \\
1 & 0 &  \cdots & 0 &0 \\
\vdots & \vdots & \ddots & \vdots & \vdots \\
0 & 0 & \cdots & 0 & 1 \\
0 & 0 & \cdots & 1 & 0
\end{bmatrix}
\end{equation}
is the nearest neighbor matrice. 

The eigen-equation of the Hamiltonian is: $\mathbf{H} \psi_\nu = E_{\nu} \psi_\nu$ where $\psi_\nu = [\mathbf{u}_\nu\;\mathbf{v}_{\nu}]^T$ with $\mathbf{u}_\nu = [\mathbf{u}_\nu^d\;\mathbf{u}_\nu^s]^T$ and $\mathbf{v}_\nu = [\mathbf{v}_\nu^d\;\mathbf{v}_\nu^s]^T$. Here, $\mathbf{u}_\nu^o$ and $\mathbf{v}_\nu^o$ represent the wave function of the electron and the hole in the QD ($o=d$) and the SC ($o=s$), respectively. Due to the definition of Majorana fermion with $\gamma_\nu^{\dagger} = \gamma_{\nu}$, it is shown that $\mathbf{u}_\nu = \mathbf{v}_\nu^*$.  Furthermore, the particle-hole symmetry inherent in the hybrid system imposes $E_\nu = 0$. The eigen-equation of $E_\nu = 0$ is
\begin{equation}
    \begin{aligned}
        \mathbf{H}_d \cdot \mathbf{u}_d + \mathbf{T}\cdot \mathbf{u}_s &= 0,\\
        \mathbf{T}^{\dagger} \cdot \mathbf{u}_d + \mathbf{H}_s \cdot \mathbf{u}_s + \mathbf{\Delta}\cdot \mathbf{u}_s^* &=0,
    \end{aligned}\label{equdus}
\end{equation}
where the index $\nu$ has been omitted for the zero-energy eigen-state. From the above two equations, it is shown that
\begin{equation}
    (\mathbf{H}_s - \mathbf{T}^{\dagger}\cdot \mathbf{H_d}\cdot \mathbf{T})\cdot \mathbf{u}_s + \mathbf{\Delta} \cdot \mathbf{u}_s^* = 0,
    \label{eqofsc}
\end{equation}
and the wave function of the electrons in QD is $\mathbf{u}_d = - \mathbf{H}_d^{-1} \cdot \mathbf{T} \cdot \mathbf{u}_s$. To solve the zero-energy wave function of the electrons in SC, we decompose $\mathbf{u}_{s}$ into its real and imaginary components: $\mathbf{u}_{s} = \mathbf{u}^{(r)}_{s} + i\mathbf{u}^{(i)}_{s}$, and they satisfy the following equations:
\begin{equation}
    (\mathbf{H}_s - \mathbf{T}^{\dagger} \cdot \mathbf{H}_d^{-1} \cdot \mathbf{T} + \lambda \mathbf{\Delta}) \cdot \mathbf{u}_s^{\lambda} = 0,\label{eqofreim}
\end{equation}
where $\lambda=-1$ and $\lambda=1$ corresponds to the real and imaginary parts of $\mathbf{u}_s$ with $\mathbf{u}_s^{-1} = \mathbf{u}_s^{i}$ and $\mathbf{u}_s^{+1} = \mathbf{u}_s^{r}$. We first consider $\mathbf{u}_s^r = [\mathbf{u}_{s,1}^r\;\dots\;\mathbf{u}_{s,n}^r \;\dots\;\mathbf{u}_{s,N}^r]^T$ with $\mathbf{u}_{s,n}^r = [u_{s,n\uparrow}^r\;u_{s,n\downarrow}^r]^T$. For $\lambda=1$, Eq. \eqref{eqofreim} can be shown as
\begin{equation}
    \begin{aligned}
        (\mu_s \sigma_0 - i\Delta_s \sigma_y) \cdot \mathbf{u}_{s,1}^r - \frac{t_s}{2} \sigma_0 \cdot \mathbf{u}_{s,2}^r - A^{\dagger} B^{-1} A \cdot\mathbf{u}_{s,1}^r &= 0,\\
        (\mu_s \sigma_0 - i\Delta_s \sigma_y) \cdot \mathbf{u}_{s,n}^r - \frac{t_s}{2} \sigma_0 \cdot (\mathbf{u}_{s,n-1}^r + \mathbf{u}_{s,n+1}^r) &= 0,\\
        (\mu_s \sigma_0 - i\Delta_s \sigma_y) \cdot \mathbf{u}_{s,N}^r - \frac{t_s}{2} \sigma_0 \cdot \mathbf{u}_{s,N-1}^r - A B^{-1} A^{\dagger} \cdot \mathbf{u}_{s,N}^r &= 0,
    \end{aligned}\label{n1N}
\end{equation}
where $n=2,\dots,N-1$, $B = {\rm{diag}}\{\mu_d + h_d,\mu_d-h_d\}$ and $A = T \sigma_0 + i\alpha \sigma_y$. We assume that the edge state exist as
\begin{equation}
    \mathbf{u}_{s,n}^r = \xi^n \mathbf{a}^r= \xi^n \begin{bmatrix}
        a_{\uparrow}^r\\
        a_{\downarrow}^r
    \end{bmatrix}
\end{equation}
with $|\xi|<1$. Then for $1<n<N$, it is shown that
\begin{equation}
    \begin{bmatrix}
        \mu_s \xi - \frac{t_s}{2} - \frac{t_s}{2} \xi^2 & -\Delta_s \xi\\
        \Delta_s \xi & \mu_s \xi - \frac{t_s}{2} - \frac{t_s}{2} \xi^2
    \end{bmatrix} \mathbf{a}^r \equiv C(\xi) \mathbf{a}^r = 0.
\end{equation}
The conditions for this equation to have a solution is ${\rm{Det}}[C(\xi)] = 0$ so that
\begin{equation}
    \xi_1 = \frac{\mu_s - i\Delta_s - \sqrt{(\mu_s - i\Delta_s)^2 - t_s^2}}{t_s},\;\xi_3 = \xi_1^{-1},\;\xi_{2(4)} = \xi_{1(3)}^*
\end{equation}
and $\mathbf{a}_l^r$ can also be solved for the given $\xi_l,\;(l=1,2)$. Therefore, the electron wave function in SC becomes $\mathbf{u}_{s,n}^r = \sum_{i=1,2} \alpha_i \xi_i^n \mathbf{a}_l^r$ with the undetermined superposition coefficient $\alpha_i$. These superposition coefficients can be determined by the first and the last equation in Eq. \eqref{n1N}, i.e. the boundary condition. And it is shown that the last equation can be satisfied naturally for $|\xi|<1$ and $N\to\infty$, then the first equation in Eq. \eqref{n1N} shows
\begin{equation}
    \sum_{i=1,2} \alpha_i \begin{bmatrix}
        (\frac{t_s}{2} - \chi_1 \xi_i) a_{i,\uparrow}^r + \sigma \xi_i a_{i,\downarrow}\\
        \sigma \xi_i a_{i,\uparrow}^r + (\frac{t_s}{2} - \chi_2 \xi_i) a_{i,\downarrow}^r
    \end{bmatrix} = 0.
\end{equation}
with 
\begin{equation}
\begin{aligned}
    \chi_1 &= [\mu_d(T^2 + \alpha^2) + h_d(\alpha^2 - T^2)]/(\mu_d^2 - h_d^2),\\
    \chi_2 &= [\mu_d(T^2 + \alpha^2) - h_d(\alpha^2 - T^2)]/(\mu_d^2 - h_d^2)
\end{aligned}
\end{equation}
and $\sigma = 2T\alpha h_d/(\mu_d^2 - h_d^2)$. Therefore, the condition for existing non-zero $\alpha_i$ gives
\begin{equation}
    \begin{aligned}
        {\rm{Det}}\left (\begin{array}{cccc}
\eta_{11} &\eta_{12}   \\
\eta_{21} &\eta_{22}  
\end{array}\right) = 0
    \end{aligned}
\end{equation}
with $\eta_{1i} = (t_s/2 - \chi_1 \xi_i) a_{i,\uparrow}^r + \sigma a_{i \downarrow}^r$ and $\eta_{2i} = (t_s/2 - \chi_2 \xi_i) a_{i,\downarrow}^r + \sigma a_{i \uparrow}^r$. This condition can be simplified as
\begin{equation}
    \left[\mu_d - \Re(\chi)\right]^2 + \Im(\chi)^2 = h_d^2 
    \label{conditions}
\end{equation}
with 
\begin{equation}
    \begin{aligned}
        \chi = \frac{2(T^2 + \alpha^2) \xi_1}{t_s}.
    \end{aligned}
\end{equation}
In the parameter region determined by Eq. \eqref{conditions}, it is shown that the imagine part, similar to the solutions of the form given in \eqref{n1N}, does not exist, i.e., $\mathbf{u}_{s,n}^i = 0$. Therefore, the first solution of the zero-energy eigen-wave is $\mathbf{u}_{1,n}^s=\mathbf{u}_{s,n}^r$. Similarly, the real and imaginary parts of the other three solutions can be solved as
\begin{equation}
\begin{aligned}
    \mathbf{u}_{2,n}^{s,i} &= \sum_{j=1,2} \beta_j \xi_j^{n} \mathbf{a}_j^i, \; \mathbf{u}_{2,n}^{s,r} =0,\\
     \mathbf{u}_{3,n}^{s,i} &= \sum_{j=3,4} \alpha_j \xi_j^{n+1-N} \mathbf{a}_j^i, \; \mathbf{u}_{3,n}^{s,r} =0,\\
     \mathbf{u}_{4,n}^{s,r} &= \sum_{j=3,4} \beta_j \xi_j^{n+1-N} \mathbf{a}_j^r,\;\mathbf{u}_{4,n}^{s,i} = 0,
\end{aligned}
\end{equation}
where $\beta_j$ and $\alpha_j$ can be determined by the boundary and normalization conditions, $\mathbf{a}_j^r\;(j=3,4)$ and $\mathbf{a}_j^i\;(j=1,2,3,4)$ can be given by $C(\xi_j) \mathbf{a}_j^r = 0\;(i=3,4) $ and $ C(\xi_j)^{\dagger} \mathbf{a}_j^i = 0\;(i=1,2,3,4)$. The three solutions presented correspond to those shown in Eq. \eqref{wave function of SC}. Specifically, they are: $\mathbf{u}_{2,n}^{s} = i \mathbf{u}_{2,n}^{s,i}$, $\mathbf{u}_{3,n}^{s} = i \mathbf{u}_{3,n}^{s,i}$ and $\mathbf{u}_{4,n}^{s} = \mathbf{u}_{4,n}^{s,r}$. Therefore, the condition for the existence of the four DPMMs, are given by Eq. \eqref{conditions} [also see Eq. \eqref{sweet spot} in the main text]. 

\twocolumngrid
\bibliography{main}
\end{document}